\documentclass[aps,prb,superscriptaddress,11pt]{revtex4-2}

\usepackage{amsmath}
\usepackage{amsfonts}
\usepackage{amssymb}
\usepackage{makeidx}
\usepackage{graphicx}
\usepackage{hyperref}
\usepackage[capitalize]{cleveref} 
\usepackage{bm} 
\usepackage{physics} 
\usepackage[linesnumbered,ruled,vlined]{algorithm2e}
\usepackage{qcircuit} 
\DeclareMathAlphabet{\bbold}{U}{bbold}{m}{n}

\usepackage[utf8]{inputenc}
\usepackage{setspace}
\usepackage{color}
\usepackage{orcidlink}

\usepackage{booktabs}
\usepackage{multirow}

\begin{document}

\title{Image Denoising via Quantum Reservoir Computing}

\author{Soumyadip Das}
\affiliation{Centre for Quantum Information, Simulation and Algorithms, School of Physics, Mathematics and Computing, The University of Western Australia, Perth, Australia}

\author{Luke Antoncich\,\orcidlink{0009-0004-9124-2650}}
\affiliation{Centre for Quantum Information, Simulation and Algorithms, School of Physics, Mathematics and Computing, The University of Western Australia, Perth, Australia}

\author{Jingbo B. Wang\,\orcidlink{0000-0001-7544-0084}}
\email{jingbo.wang@uwa.edu.au}
\affiliation{Centre for Quantum Information, Simulation and Algorithms, School of Physics, Mathematics and Computing, The University of Western Australia, Perth, Australia}

\begin{abstract}
Quantum Reservoir Computing (QRC) leverages the natural dynamics of quantum systems for information processing, without requiring a fault-tolerant quantum computer. In this work, we apply QRC within a hybrid quantum–classical framework for image denoising. The quantum reservoir is implemented using a Rydberg atom array, while a classical neural network serves as the readout layer. To prepare the input, images are first compressed using Principal Component Analysis (PCA), reducing their dimensionality to match the size of the atom array. Each feature vector is encoded into local detuning parameters of a time-dependent Hamiltonian governing the Rydberg system. As the system evolves, it generates nonlinear embeddings through the measurement of observables across multiple time steps. These temporal embeddings capture complex correlations, which are fed into a classical neural network to reconstruct the denoised images. To evaluate performance, we compare this QRC-assisted model against a baseline architecture consisting of PCA followed by a dense neural network, trained under identical conditions. Our results show that the QRC-based approach achieves improved image sharpness and similar structural recovery compared to the PCA-based model. We demonstrate the practical viability of this framework through experiments on QuEra's Aquila neutral-atom processor, leveraging its programmable atom arrays to physically realize the reservoir dynamics. 

\end{abstract}

\maketitle

\section{Introduction}

Image denoising is a core problem in computer vision and signal processing, with the aim of recovering clean images from noisy observations. Classical techniques, such as Principal Component Analysis (PCA) \cite{jolliffe2002principal} and convolutional neural networks (CNNs) \cite{zhang2017beyond}, have been widely adopted due to their effectiveness in capturing statistical correlations and spatial patterns. PCA reduces dimensionality by identifying principal components that explain the variance in the data, while CNNs learn hierarchical features through large-scale training on labeled datasets. Despite their success, these methods often face limitations when dealing with structured noise, dynamic image content, or nonlinear dependencies that evolve over time \cite{Zhao_2019}. In such cases, the assumptions underlying classical models, such as linearity or stationarity, may not hold, leading to suboptimal denoising performance. 

Quantum computing offers powerful new tools for high-dimensional and nonlinear information processing with the potential to outperform classical methods in certain tasks \cite{biamonte2017quantum, Chiew2019, Yusen2023}. In recent years, several quantum approaches have  been explored for image denoising \cite{kerger2023quantumimagedenoisingframework, zhou2024imagedenoisingmachinelearning, 2020IJTP...59.3348C}, demonstrating promising results in noise suppression and feature recovery. One potential quantum approach is Quantum Reservoir Computing (QRC) \cite{fujii2017harnessing, nakajima2020quantum}, which utilizes the complex dynamics of quantum many-body systems as computational reservoirs. In QRC, input data is encoded in the control parameters of a quantum system, typically through a time-dependent Hamiltonian, which drives the evolution of the system. This evolution naturally maps the input to a high-dimensional feature space, due to the inherent complexity of the system. Outputs are extracted from measured observables over time and processed using a straight-forward readout.

In this study, we apply QRC to image denoising by encoding image features as inputs to a Rydberg-atom reservoir, where Hamiltonian dynamics processes the noisy data. A linear readout reconstructs the cleaned images, and performance is benchmarked against a PCA-based classical baseline. The QRC model achieves higher image sharpness and similar structural similarity, demonstrating the utility of quantum correlations to extract information from noise. These findings support the potential of analog quantum systems as computational platforms for signal recovery and pattern recognition. They also demonstrate how controlled driving fields and detuning pulses can enhance nonlinear processing without requiring full quantum error correction. 

\section{Methodology}
\label{sec:classical_heuristics}

\subsection{Overview}
The core of this work is a novel hybrid quantum-classical framework designed for image denoising. The central idea is that the complex, high-dimensional dynamics of a quantum many-body system can serve as a powerful computational resource for feature extraction. Our proposed framework integrates a Quantum Reservoir Computer (QRC)~\cite{kornjača2024largescalequantumreservoirlearning,nakajima2021reservoir,kutvonen2020reservoir,nakajima2020quantum} with a classical all-to-all neural network.

The methodology unfolds in three primary stages:
\begin{enumerate}
    \item \textbf{Preprocessing:} Noisy input images are first subjected to Principal Component Analysis (PCA)\cite{jolliffe2002principal} for dimensionality reduction. This compresses the high-dimensional image data into a low-dimensional representation suitable for encoding onto a near-term quantum device.
    \item \textbf{Quantum Feature Mapping:} This representation is used to drive the dynamics of a quantum reservoir. The time evolution of the quantum system, governed by the Hamiltonian, nonlinearly maps the input features into the system's Hilbert space.
    \item \textbf{Classical Reconstruction:} The final state of the quantum reservoir is measured to extract a classical feature vector. This vector, which now encodes complex correlations from the input data, is fed into a classical feedforward neural network (the "readout network") that is trained to reconstruct the clean image.
\end{enumerate}

\begin{figure*}[hbt!]
    \centering
    \includegraphics[width=\linewidth]{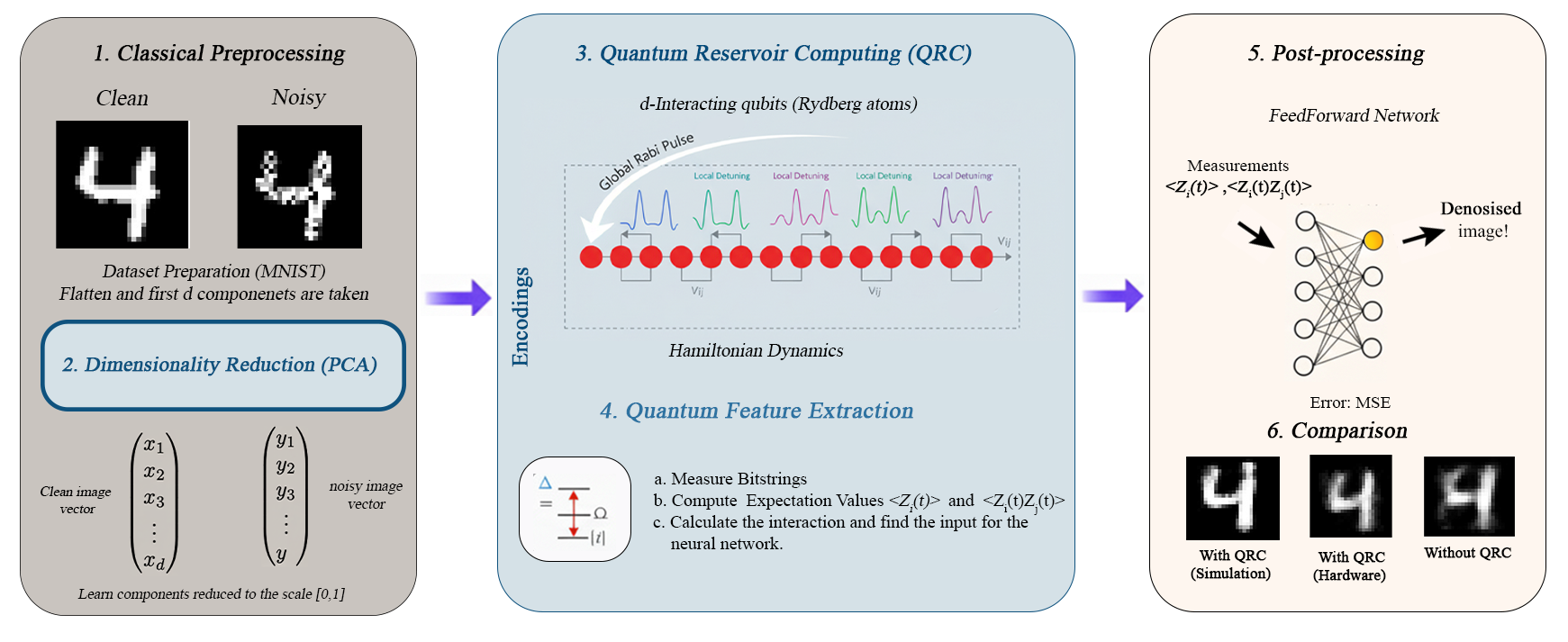}
    \caption{\textbf{Schematic overview of the hybrid quantum–classical image denoising scheme based on Quantum Reservoir Computing (QRC).} 
Noisy MNIST images undergo classical preprocessing (flattening and normalization) and PCA dimensionality reduction to \textit{d} components (scaled to [0,1]), which are encoded as local detunings in a chain of \textit{d} interacting Rydberg atoms. The quantum reservoir evolves under time-varying global Rabi pulses \(\Omega(t)\) and detunings, and features are extracted from expectation values of single-site \(\langle Z_i(t) \rangle\) and pairwise \(\langle Z_i(t) Z_j(t) \rangle\) observables across multiple timesteps. A classical feedforward neural network reconstructs the clean image from these quantum embeddings using an MSE training objective. Compared to a classical PCA baseline, the QRC approach consistently produces sharper images while maintaining near-identical structure, with associated metrics increasing as the reservoir size increases up to the maximum simulated dimension of \textit{d}=18 atoms.
}
    \label{fig:Overview}
\end{figure*}

For evaluation, we benchmark our hybrid model against a purely classical baseline that uses the same PCA features but feeds them directly into an identical readout network, helping isolate the performance contribution of the quantum reservoir. Figure~\ref{fig:Overview} provides a schematic overview of the hybrid quantum–classical image denoising framework. 

\subsection{Dataset and Preprocessing}
Let the dataset be denoted by $\mathcal{D} = \{(X_i, Y_i)\}_{i=1}^{N}$, a collection of $N$ image pairs. For each pair, $X_i \in [0,1]^{H \times W}$ is the ground-truth clean grayscale image, and $Y_i \in [0,1]^{H \times W}$ is its noisy counterpart. The noise is introduced via the model:
\begin{equation}
    Y_i = \mathcal{N}(X_i) = X_i \odot (1 + \epsilon_i),
\end{equation}
where $\odot$ signifies the element-wise Hadamard product. The noise term $\epsilon_i$ is a matrix of random variables drawn from a zero-mean distribution, such as a Gaussian distribution $\mathcal{N}(0, \sigma^2)$, modeling a combination of additive and multiplicative noise often seen in real-world imaging systems.

For processing, each image is unrolled into a vector, $\mathbf{x}_i = \mathrm{vec}(X_i) \in \mathbb{R}^M$, where $M = H \times W$ is the total number of pixels. This allows the image to be treated as a single data point in a high-dimensional space. The resulting vectors $\{\mathbf{x}_i, \mathbf{y}_i\}$ are normalized to ensure that all pixel values are within the $[0,1]$ range.

\subsection{Dimensionality Reduction via PCA}
Given the practical limitations of current quantum hardware (i.e., the number of available qubits, or 'sites'), it is infeasible to directly encode an entire image vector $\mathbf{y}_i \in \mathbb{R}^M$ onto the quantum device. We therefore employ Principal Component Analysis (PCA) as an unsupervised linear dimensionality reduction technique.

PCA identifies an optimal orthogonal basis $\{\mathbf{u}_1, \dots, \mathbf{u}_M\}$ for the clean training data by finding the directions of maximal variance. The transformation projects the high-dimensional noisy vector $\mathbf{y}_i$ onto the subspace spanned by the first $d \ll M$ principal components:
\begin{equation}
    \mathbf{z}_i = U_d^\top (\mathbf{y}_i - \bar{\mathbf{y}}).
\end{equation}
Here, $U_d = [\mathbf{u}_1, \dots, \mathbf{u}_d] \in \mathbb{R}^{M \times d}$ is the projection matrix containing the top $d$ eigenvectors of the data's covariance matrix, and $\bar{\mathbf{y}}$ is the mean of the noisy training vectors. The hyperparameter $d$ is chosen to retain a significant portion of the total variance in the original data. The resulting low-dimensional vector $\mathbf{z}_i \in \mathbb{R}^d$ serves as a compressed representation of the noisy image, which is now compact enough to be encoded into the quantum reservoir.

\subsection{Quantum Reservoir Encoding}
The quantum reservoir is the computational core of our feature mapping process. It consists of an interacting chain of $d$ qubits, where each qubit is associated with one component of the PCA-reduced input vector $\mathbf{z}_i$. The system's dynamics are governed by a time-dependent transverse-field Ising Hamiltonian, a standard model for interacting spin systems:
\begin{equation}
    H(t) = \sum_{i=1}^{d} \big[ \Omega_i(t) \sigma_x^{(i)} + \Delta_i(t) n_i \big] + \sum_{i<j} V_{ij} n_i n_j .
\end{equation}
The terms in the Hamiltonian have clear physical interpretations:
\begin{itemize}
    \item $\Omega_i(t) \sigma_x^{(i)}$: The Rabi frequency $\Omega_i(t)$ drives coherent transitions between the ground ($\ket{0}$) and excited ($\ket{1}$) states of qubit $i$, governed by the Pauli-X operator $\sigma_x^{(i)}$.
    \item $\Delta_i(t) n_i$: The detuning $\Delta_i(t)$ controls the energy offset of the excited state of qubit $i$, where $n_i = \frac{1}{2}(1 + \sigma_z^{(i)})$ is the number operator. This term is used to encode the input data.
    \item $V_{ij} n_i n_j$: The interaction term $V_{ij}$ describes the coupling strength between qubits $i$ and $j$. This term generates entanglement and complex correlations within the reservoir.
\end{itemize}

The input vector $\mathbf{z}_i = (z_{i,1}, \dots, z_{i,d})^\top$ is encoded into the reservoir's dynamics by modulating the local detuning profiles. A simple linear encoding scheme is used:
\begin{equation}
    \Delta_i(t) = \Delta_0 + k \cdot z_{i,i},
\end{equation}
where $\Delta_0$ is a constant baseline detuning and $k$ is a scaling factor. The reservoir is initialized in the ground state $\ket{\psi(0)} = \ket{0}^{\otimes d}$ and evolves for a total time $T$ according to the time-dependent Schrödinger equation:
\begin{equation}
    i\hbar \frac{d}{dt} \ket{\psi(t)} = H(t) \ket{\psi(t)}.
\end{equation}
This evolution is simulated numerically, yielding the final state $\ket{\psi(T)}$. This state is a highly complex, nonlinear function of the input vector $\mathbf{z}_i$.

The quantum reservoir was driven using simple time-independent control fields. 
First, the Rabi drive was applied as a constant pulse,
\[
\Omega(t) = 2\pi\ \text{rad}/\mu\text{s}.
\]

\noindent The encoding scale used for the detuning terms was set to $\Delta = 9.0\ \text{rad}/\mu\text{s}$. The global detuning was held constant at half this value,
\[
\Delta_{\text{global}} = \frac{\Delta}{2} = 4.5\ \text{rad}/\mu\text{s}.
\]

\noindent Each atom was also assigned a constant local detuning proportional to its encoded input value. Let $x_{\mathrm{vec}} = (x_1, x_2, \ldots, x_N)$ denote the scaled PCA components mapped to the sites.  
The local detuning profile was then
\[
\Delta_{\text{local},i} = -\Delta\, x_i = -9.0\, x_i\ \text{rad}/\mu\text{s},
\]
which provides a fixed, site-dependent energy shift for each atom.

\subsection{Quantum Feature Extraction}
To translate the information stored in the evolved quantum state $\ket{\psi(T)}$ into a classical format, we perform measurements \cite{schuld2019qmlhilbert}. The system's state is sampled at $L$ discrete time points $\{t_1, t_2, \dots, t_L = T\}$ during its evolution. At each time step $t_l$, we estimate the expectation values of single-qubit and two-qubit observables in the computational basis. For each input, we perform $K$ repeated state preparations and measurements ('shots') to gather statistics.

The expectation values are estimated as follows:
\begin{itemize}
    \item Single-qubit observables: $\langle Z_i \rangle_{t_l} = \frac{1}{K} \sum_{k=1}^{K} (-1)^{b_{k,i}^{(l)}}$, which captures local information about each qubit.
    \item Two-qubit correlation observables: $\langle Z_i Z_j \rangle_{t_l} = \frac{1}{K} \sum_{k=1}^{K} (-1)^{b_{k,i}^{(l)} \oplus b_{k,j}^{(l)}}$, which captures the correlations and entanglement generated between pairs of qubits by the $V_{ij}$ term. 
\end{itemize}

\noindent Here, $b_{k,i}^{(l)}$ is the binary outcome for qubit $i$ in shot $k$ at time $t_l$, and $\oplus$ is addition modulo 2 (XOR).

The final quantum feature embedding $\mathbf{r}_i \in \mathbb{R}^R$ for the input $\mathbf{y}_i$ is constructed by concatenating all these measured expectation values across all relevant sites and time steps:
\begin{equation}
    \mathbf{r}_i = \mathrm{concat}\big( \{\langle Z_p \rangle_{t_l}, \langle Z_p Z_q \rangle_{t_l} \}_{p<q, l=1..L} \big).
\end{equation}
The dimensionality of this feature vector is $R = L \times (d + \frac{d(d-1)}{2})$. This vector serves as a fixed, high-dimensional, and non-linear representation of the original input, which is now ready to be interpreted by a classical machine learning model.

\subsection{Classical Readout Network}
The final step is to map the quantum feature embedding $\mathbf{r}_i$ back to a denoised image. This is accomplished by a trainable classical readout network, specifically a Multi-Layer Perceptron (MLP). This network defines a parametric function $f_\theta$ that learns the mapping:
\begin{equation}
    \hat{\mathbf{x}}_i = f_\theta(\mathbf{r}_i),
\end{equation}
where $\theta$ represents the set of all trainable weights and biases in the network. The MLP consists of an input layer of size $R$, several hidden layers with a non-linear activation function (e.g., ReLU), and an output layer of size $M$ with a Sigmoid activation function to ensure the output pixel values are constrained to $[0,1]$.

The network is trained end-to-end by minimizing the Mean Squared Error (MSE) loss function between the reconstructed image $\hat{\mathbf{x}}_i$ and the ground-truth clean image $\mathbf{x}_i$:
\begin{equation}
    \mathcal{L}(\theta) = \frac{1}{N} \sum_{i=1}^{N} \| f_\theta(\mathbf{r}_i) - \mathbf{x}_i \|_2^2~.
\end{equation}
Optimization of the parameters $\theta$ is performed using the Adam optimizer, a stochastic gradient-based algorithm, via standard backpropagation. Only the classical readout network's parameters $\theta$ are trained; the quantum reservoir's dynamics are fixed.

\subsection{Classical Baseline Model}
To properly assess the advantage conferred by the quantum feature mapping, a purely classical baseline model is established as a control. This baseline replaces the entire quantum processing pipeline with a direct connection from the PCA stage to the readout network. The baseline network, $f_{\phi}$, thus learns the mapping:
\begin{equation}
    \hat{\mathbf{x}}_i^{(\mathrm{PCA})} = f_{\phi}(\mathbf{z}_i).
\end{equation}
To ensure a fair comparison, the architecture of this baseline network (number of layers, neurons, activation functions) is identical to the one used in the hybrid model. It is also trained using the same loss function, optimizer, and training procedure.

\subsection{Evaluation Metrics}
Model performance is quantitatively assessed using three standard and complementary metrics from the field of image processing:
\begin{itemize}
    \item Mean Squared Error (MSE): Measures the average squared difference between the pixels of the reconstructed and ground-truth images. It provides a direct measure of pixel-wise reconstruction fidelity:
    \begin{equation}
        \mathrm{MSE} = \frac{1}{M} \sum_{j=1}^{M} (x_{ij} - \hat{x}_{ij})^2.
    \end{equation}
    \item Tenegrad Sharpness (TENG): A focus and edge-sharpness metric based on the magnitude of image gradients computed via the Sobel operator. Higher Tenegrad values indicate stronger edge responses and therefore sharper reconstructions:
    \begin{equation}
        \mathrm{TENG}(X) = \frac{1}{M} \sum_{j=1}^{M} 
        \mathbb{1}\!\left(G_j > T\right)\, G_j^2,
    \end{equation}
    where $G_j$ is the Sobel gradient magnitude at pixel $j$, $T$ is a threshold that suppresses noise-dominated gradients, and $\mathbb{1}(\cdot)$ is the indicator function. The metric effectively captures local edge strength rather than global pixel-wise error.
    \item Structural Similarity Index (SSIM) \cite{inproceedings}: A perceptual metric that assesses the similarity between two images based on luminance, contrast, and structure, which aligns more closely with human visual perception than pixel-wise metrics like MSE:
    \begin{equation}
        \mathrm{SSIM}(X, \hat{X}) = 
        \frac{(2\mu_X \mu_{\hat{X}} + c_1)(2\sigma_{X\hat{X}} + c_2)}%
        {(\mu_X^2 + \mu_{\hat{X}}^2 + c_1)(\sigma_X^2 + \sigma_{\hat{X}}^2 + c_2)}.
    \end{equation}
\end{itemize}

\subsection{Overall Framework}
In summary, the proposed hybrid denoising pipeline implements the following compositional mapping from a noisy image $\mathbf{y}_i$ to its reconstructed clean version $\hat{\mathbf{x}}_i$:
\begin{equation}
    \hat{\mathbf{x}}_i = (f_\theta \circ g_Q \circ h_\mathrm{PCA}) (\mathbf{y}_i),
\end{equation}
where $h_\mathrm{PCA}$ is the linear PCA projection, $g_Q$ represents the non-trainable, non-linear quantum feature mapping, and $f_\theta$ is the trainable classical reconstruction network. The baseline model simply omits the $g_Q$ step. This structure leverages the quantum system's natural capacity for generating complex correlations to create a powerful feature expansion, enabling the classical network to solve the denoising problem more effectively than it could with linearly-compressed features alone.

\section{Results and Dicussion}

To evaluate the performance of our Quantum Reservoir Computing (QRC) based image denoiser, we performed experiments on the \cite{Deng2012TheMD} MNIST dataset, a widely adopted benchmark comprising 60,000 training and 10,000 test images of 28×28 grayscale handwritten digits range between 0 to 9. For the setup, we selected the first 1000 training images and 200 test images to manage computational overhead during emulations with Bloqade \cite{Wurtz:2023lrt}. Noisy datasets were synthesized by applying multiplicative speckle noise, modeled as \( I{\tilde{}} = I \cdot (1 + \sigma \cdot \mathcal{N}(0,1)) \), with \(\sigma = 0.7\), cut to [0,1] to simulate realistic distortions encountered in optical imaging and sensor noise. This noise type was chosen for its multiplicative nature, which preserves relative intensities but introduces spatially varying artifacts, posing a challenging denoising scenario.

The quantity of PCA components and, consequently, the quantity of Rydberg atoms in the QRC chain was altered from 4 to 18 atoms to examine scalability and performance balances. Noisy images were applied to these PCA bases, normalized to [0,1] for each component, and represented as local detunings within a one-dimensional chain structure with a lattice spacing of 10 $\mu$m. The system progressed through 8 timesteps (dt=0.5 $\mu$s), utilizing 1000 shots for each emulation to calculate expectation values of individual-site \(Z_i\) and pairwise \(Z_iZ_{j}\) observables, resulting in high-dimensional embeddings (e.g., dimension 288 for 8 atoms: 8 timesteps × (8 Z + 28 ZZ)).

A TensorFlow sequential model (two hidden layers: 1024 and 512 ReLU units, sigmoid output for 784 pixels) was trained on these embeddings to regress clean pixel values, using MSE loss, Adam optimizer (lr=0.001), batch size 64, and 500 epochs. A dropout rate of 0.3 was used as a form of regularization and a patience of 20 was chosen for early stopping. As a baseline, an identical model mapped noisy PCA features directly to clean images, bypassing the QRC. Performance was quantified on the 200 noisy test images which were held out from training.

\begin{figure*}[hbt!]
    \centering
    \includegraphics[width=\linewidth]{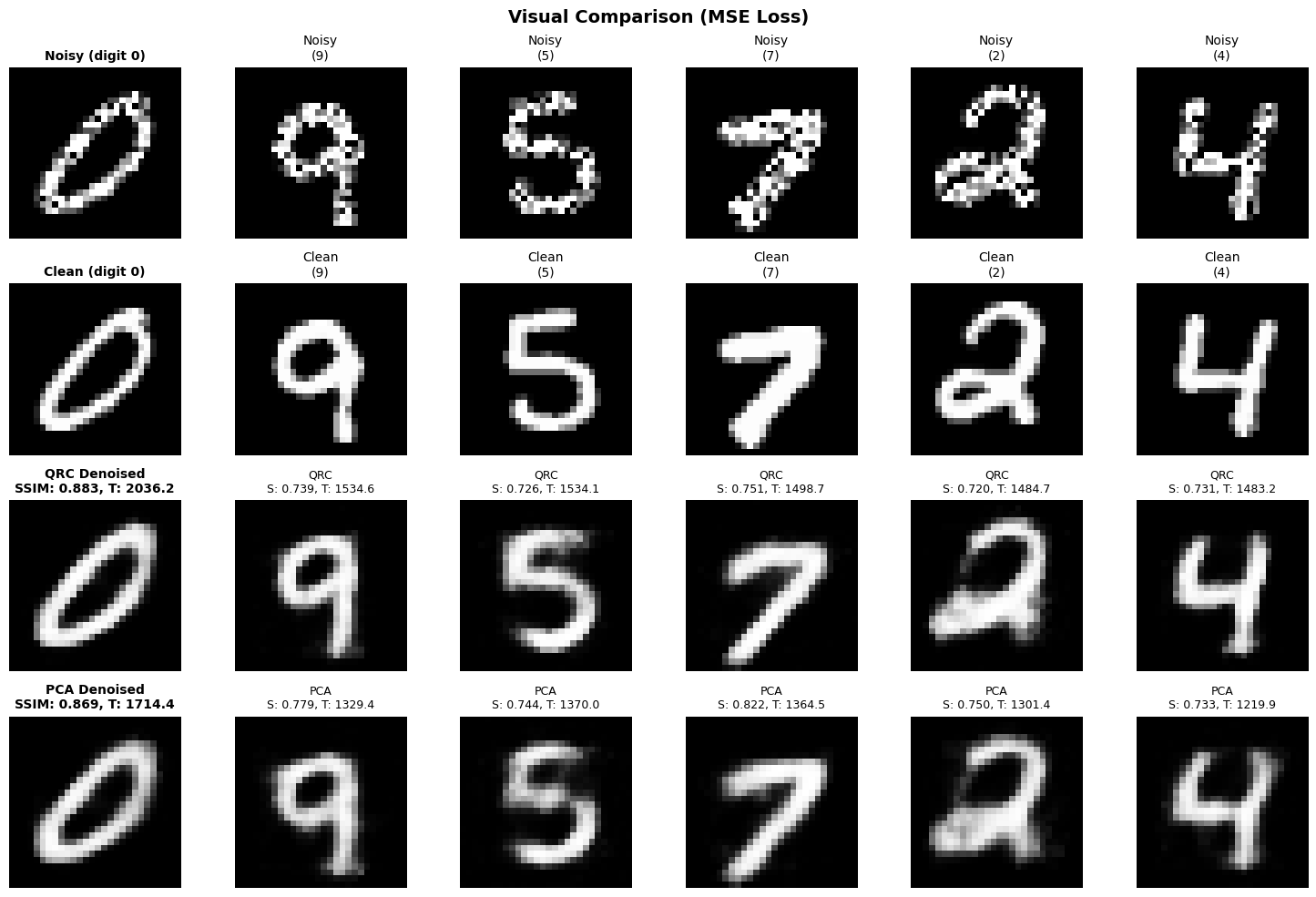}
    \caption{\textbf{Comparison of denoising performance on MNIST dataset}: The first row shows the noisy inputs, the second row shows the clean reference images, the third row shows results from Quantum Reservoir Computing (QRC) denoising using an 18-atom chain, and the fourth row shows outputs from classical PCA-based denoising. The QRC-denoised images have sharper edges than their classical counterparts, while maintaining similar structure.}
    \label{fig:comparison}
\end{figure*}

Qualitatively, visualizations of the denoised outputs showed that QRC reconstructions consistently exhibited sharper edges and fewer residual artifacts across all reservoir sizes from 4 to 18 atoms. Figure~\ref{fig:comparison} compares clean and noisy MNIST images with the reconstructions generated by 18-atom QRC and the corresponding PCA model. Both models are able to significantly reduce the added speckle noise.

\begin{figure*}[hbt]
    \centering
    \includegraphics[width=\linewidth]{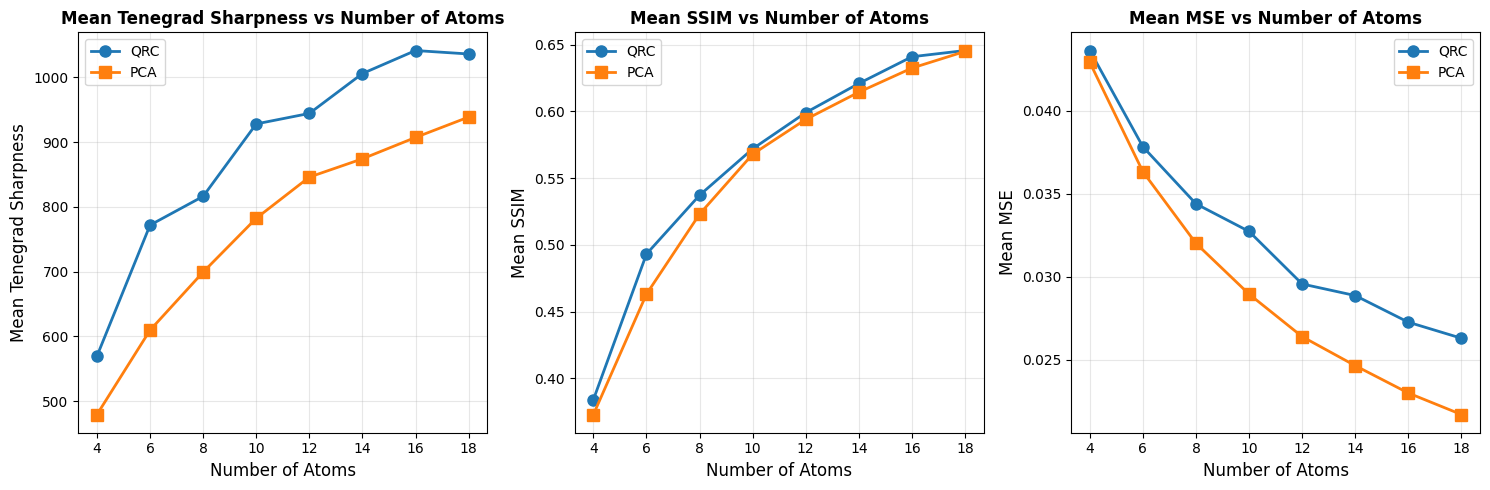}

    \caption{\textbf{Comparison between Quantum Reservoir Computing (QRC) and classical PCA-based denoising performance across different reservoir sizes}: The QRC model achieves a higher Tenegrad Sharpness (TENG) and slightly better Structural Similarity Index (SSIM) than the PCA model, while the PCA model achieves a lower (better) MSE. As the atom count increases both models continue to improve, with some evidence of saturation at the highest count. These results indicate the potential of QRC as a tool to denoise images and restore their sharpness, which has applications in medical contexts for example.}
    \label{fig:results}
\end{figure*}

The analysis across all tested reservoir sizes is summarized  in Figure~\ref{fig:results} and Table~\ref{tab:metrics}. In terms of SSIM, QRC matched PCA across all atom counts, indicating that the quantum reservoir preserves global structural features to a similar degree despite its nonlinear dynamics. QRC showed a clear advantage in TENG, outperforming PCA in every configuration. These higher sharpness scores suggest that the reservoir’s nonlinear evolution amplifies edge-relevant features, leading to reconstructions with stronger local gradients than those obtained from linear PCA. Conversely, QRC exhibited higher MSE across all settings, reflecting greater pixel-level variability. This elevated error likely arises from the stochasticity and nonlinear spreading of information within the reservoir, which can introduce small local deviations even when overall structure and edge strength are preserved.

\setlength{\tabcolsep}{6pt}  
\begin{table}[hbt]
\centering
\caption{Performance comparison across PCA dimensions. Values are mean $\pm$ standard deviation; best value per row for each metric is shown in bold.}
\label{tab:metrics}
\small
\begin{tabular}{l|cc|cc|cc}
\toprule
\multirow{2}{*}{\textbf{Atoms}} &
\multicolumn{2}{c}{\textbf{MSE}} &
\multicolumn{2}{c}{\textbf{SSIM}} &\multicolumn{2}{c}{\textbf{TENG}} \\
\cmidrule(lr){2-3} \cmidrule(lr){4-5} \cmidrule(lr){6-7}
 & \textbf{QRC} & \textbf{PCA} & \textbf{QRC} & \textbf{PCA} & \textbf{QRC} & \textbf{PCA} \\
\midrule
4  & 0.044$\pm$0.018 & \textbf{0.043$\pm$0.015} & \textbf{0.384$\pm$0.149} & 0.373$\pm$0.131 & \textbf{569.7$\pm$245.2}  & 479.4$\pm$236.2 \\
6  & 0.038$\pm$0.017 & \textbf{0.036$\pm$0.015} & \textbf{0.493$\pm$0.171} & 0.463$\pm$0.144 & \textbf{771.9$\pm$290.3}  & 609.7$\pm$284.6 \\
8  & 0.034$\pm$0.016 & \textbf{0.032$\pm$0.014} & \textbf{0.537$\pm$0.169} & 0.523$\pm$0.148 & \textbf{816.3$\pm$306.0}  & 699.3$\pm$297.9 \\
10 & 0.033$\pm$0.017 & \textbf{0.029$\pm$0.013} & \textbf{0.572$\pm$0.156} & 0.568$\pm$0.139 & \textbf{928.1$\pm$304.7}  & 782.4$\pm$303.6 \\
12 & 0.030$\pm$0.015 & \textbf{0.026$\pm$0.012} & \textbf{0.599$\pm$0.154} & 0.594$\pm$0.127 & \textbf{944.3$\pm$304.3}  & 846.0$\pm$309.8 \\
14 & 0.029$\pm$0.015 & \textbf{0.025$\pm$0.011} & \textbf{0.621$\pm$0.151} & 0.614$\pm$0.123 & \textbf{1005.8$\pm$311.9}  & 874.2$\pm$313.9 \\
16 & 0.027$\pm$0.014 & \textbf{0.023$\pm$0.010} & \textbf{0.641$\pm$0.145} & 0.632$\pm$0.124 & \textbf{1041.4$\pm$315.0} & 907.2$\pm$322.9 \\
18 & 0.026$\pm$0.013 & \textbf{0.022$\pm$0.010} & \textbf{0.646$\pm$0.146} & 0.645$\pm$0.123 & \textbf{1036.0$\pm$315.8} & 938.6$\pm$327.9 \\
\bottomrule
\end{tabular}
\end{table}

\begin{figure}
    \centering
    \includegraphics[width=0.3\linewidth]{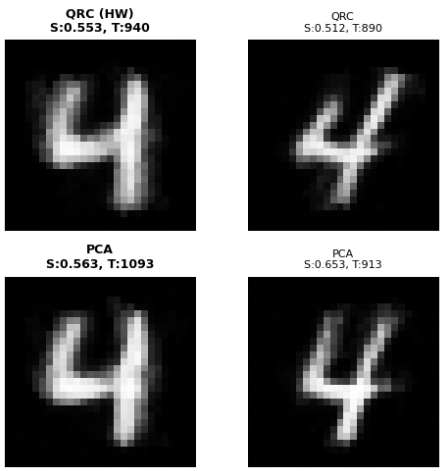}
    \caption{\textbf{Comparison of denoising performance between QRC embeddings generated by hardware and the PCA-based model}: The first row shows results from Quantum Reservoir Computing (QRC) denoising using a 14-atom chain on the hardware device Aquila, and the second row shows outputs from classical PCA-based denoising. The QRC-denoised images have more or less similar sharpness to those generated by the PCA model.}
    \label{fig:hardware-comparison}
\end{figure}

Experimental execution was performed on QuEra's Aquila neutral-atom processor. Due to the significant computational overhead of sampling observables at various time evolutions for each input, the study focused on 100 training and 20 testing instances. This provided a representative proof-of-concept while managing the total quantum resources required for high-resolution temporal tracking. 
 With 14-atom chains and sampling the reservoir in 5 time steps, the QRC model trained with hardware embeddings did not perform as well as in the emulations. This is likely a combination of inherent noise in the system that is not present in emulations, as well as the restricted shot count (200 shots with hardware as opposed 1000 emulation shots). For some data samples the hardware results were able to achieve similar results to the PCA model, which can be seen in Figure~\ref{fig:hardware-comparison}. With the potential to take more shots and with longer coherence times for hardware atoms, it is feasible that this technique could be implemented with near-term quantum devices and with larger atom counts.

\section{conclusion}
\label{sec:conclusion}

This study demonstrates that Quantum Reservoir Computing offers a viable approach to image denoising. Across all reservoir sizes tested, QRC produced reconstructions with  higher Tenegrad Sharpness (TENG) and similar Structural Similarity Index Measure (SSIM) values compared to PCA, indicating stronger edge recovery without compromising global structural fidelity. Sharper reconstructions are especially relevant for downstream tasks where fine-scale gradients influence diagnostic or classification accuracy, such as medical imaging, pattern recognition, and feature extraction.

A notable finding is that QRC performance continued to improve with increasing atom number, with slight evidence of saturation at 18 atoms. This scaling behaviour suggests that larger quantum reservoirs may unlock even richer nonlinear feature transformations. A promising direction for future work is to explore the extent of this trend on larger neutral-atom devices and to assess whether similar gains persist for higher-resolution datasets or more challenging noise models. Overall, these results highlight the potential of quantum dynamical embeddings to enhance perceptual image quality and motivate further investigation into the capabilities of large-scale quantum reservoirs for vision and signal-processing applications.

In conclusion, QRC offers a promising direction for energy-efficient, hardware-native signal processing in image denoising. By leveraging the high-dimensional structure of Hilbert space and the complex intrinsic dynamics of quantum systems, QRC may separate signal from noise more effectively than classical reservoirs. This design requires only a simple classical readout layer and avoids the computational cost of backpropagation, making QRC particularly suited for small-data and resource-constrained environments. 
Future work will focus on rigorous benchmarking against classical and hybrid methods, scaling strategies, and assessing performance on real-world datasets.

\section*{Acknowledgements}
This research was supported by the Critical Technologies Challenge Program of the Australian Government Department of Industry, Science and Resources. LA gratefully acknowledges funding from the Australian Government Research Training Program scholarship. Numerical simulations were conducted on Setonix, provided by the Pawsey Supercomputing Research Centre. Access to the QuEra Aquila processor was also facilitated through Pawsey Supercomputing Research Centre. The authors acknowledge valuable discussions with Tommaso Macri and Jonathan Wurtz from QuEra. 

\bibliography{refs}

\end{document}